\documentclass[showpacs,amsmath,amssymb,aps,pra,twocolumn]{revtex4}
\usepackage{bm}
\usepackage{amsfonts}
\usepackage{amssymb}
\usepackage{graphicx}
\begin{document}

\title{Electromagnetic radiation by gravitating bodies}
\author{Iwo Bialynicki-Birula}\email{birula@cft.edu.pl}
\affiliation{Center for Theoretical Physics, Polish Academy of Sciences\\
Al. Lotnik\'ow 32/46, 02-668 Warsaw, Poland and\\ Institute of Theoretical Physics, University of Warsaw, Warsaw, Poland}
\author{Zofia Bialynicka-Birula}\affiliation{Institute of Physics, Polish Academy of
Sciences\\
Al. Lotnik\'ow 32/46, 02-668 Warsaw, Poland}

\begin{abstract}
Gravitating bodies in motion, regardless of their constitution, always produce electromagnetic radiation in the form of photon pairs. This phenomenon is an analog of the radiation caused by the motion of dielectric (or magnetic) bodies. It is a member of a wide class of phenomena named dynamical Casimir effects, and it may be viewed as the squeezing of the electromagnetic vacuum. Production of photon pairs is a purely quantum-mechanical effect. Unfortunately, as we show, the emitted radiation is extremely weak as compared to radiation produced by other mechanisms.
\end{abstract}
\maketitle
\section{Introduction}

The aim of this paper is to calculate the probability of photon emission by gravitating orbiting bodies. There are two separate mechanisms responsible for this process: electromagnetic and gravitational. The electromagnetic mechanism has its source in the variation of the material coefficients with time. Such changes may be caused, for example, by electromagnetic waves impinging on a nonlinear medium or by mechanical vibrations. The electromagnetic production of photon pairs by time-varying media and its connection with squeezing were analyzed by us in the general case some time ago \cite{osa}. A closely related problem is that of oscillating cavity walls (mirrors). This problem, named the dynamical Casimir effect, was investigated for the first time by Moore \cite{moore}.  A thorough review paper on the dynamical Casimir effect by Dodonov \cite{dod} contains a long list of references. Later this line of research was extended to include all kinds of nonadiabatic transitions such as, for example, a proposal by Scully {\em et al.} \cite{scully} to generate photons by accelerating a neutral atom in a resonant cavity.

Gravitational particle production due to changes of the metric with time was predicted a long time ago by Schr\"odinger \cite{schr} and later pursued by many others (cf., for example, \cite{ima,parker,unruh,bd,ford}). This effect has been considered mostly in the cosmological context --- the time-dependent metric was usually taken in the Friedman-Robertson-Walker form. In contrast, we consider here only local perturbations of the metric, avoiding global cosmological issues.

In the case of gravitating orbiting bodies with total mass $M$ and angular frequency of the orbital motion $\Omega$, there are two independent dimensional parameters: say $GM\Omega/c^3$ and $\hbar c/GM^2$. Therefore it is not possible to estimate the order of magnitude of the effect by dimensional analysis (it could be any function of these two parameters) and a detailed calculation is necessary.

The starting point of our investigation is the set of time-dependent Maxwell equations,
\begin{align}\label{max}
\partial_t{\bm B}({\bm r}, t)=-\nabla\!\times\!{\bm E}({\bm r}, t),\;
\partial_t{\bm D}({\bm r}, t)=\nabla\!\times\!{\bm H}({\bm r}, t),
\end{align}
which are valid in classical as well as in quantum electrodynamics. For space-time-dependent $\epsilon$ and $\mu$ they can be written in the form
\begin{subequations}\label{max1}
\begin{align}
\partial_t{\bm B}({\bm r}, t)+\nabla\!\times\!\frac{{\bm D}({\bm r},t)}{\epsilon_0}
&=\nabla\!\times\!\alpha({\bm r}, t)\frac{{\bm D}({\bm r},t)}{\epsilon_0},\\
\partial_t{\bm D}({\bm r}, t)-\nabla\!\times\!\frac{{\bm B}({\bm r}, t)}{\mu_0}
&=-\nabla\!\times\!\beta({\bm r}, t)\frac{{\bm B}({\bm r}, t)}{\mu_0},
\end{align}
\end{subequations}
where
\begin{align}
\alpha({\bm r}, t)=\left(\frac{\epsilon_0}{\epsilon({\bm r}, t)}-1\right),\;\;
\beta({\bm r}, t)=\left(\frac{\mu_0}{\mu({\bm r}, t)}-1\right).
\end{align}
The right-hand sides in Eqs.~(\ref{max1}) are responsible for the creation of photons. This process is best described with the help of the $S$ operator (given as a time-ordered exponential),
\begin{align}
S&=T\exp\left(-\frac{i}{\hbar}\int\!d^3r\,dt\,H_I({\bm r}, t)\right),\\
H_I({\bm r}, t)&=\frac{\alpha({\bm r}, t)}{2\epsilon_0}{\bm D}^2({\bm r}, t)
+\frac{\beta({\bm r}, t)}{2\mu_0}{\bm B}^2({\bm r}, t),\label{hint}
\end{align}
where ${\bm D}({\bm r}, t)$ and ${\bm B}({\bm r}, t)$ are the operators of the free field. In what follows we shall use the Riemann-Silberstein vector \cite{pwf,qed},
\begin{align}
&{\bm F}({\bm r}, t)=\frac{{\bm D}({\bm r}, t)}{\sqrt{2\epsilon_0}}+i\frac{{\bm B}({\bm r}, t)}{\sqrt{2\mu_0}}=\int\!d^3k\sqrt{\frac{\hbar\omega}{(2\pi)^3}}\nonumber\\
&\times{\bm e}({\bm k})\left(a_L({\bm k})e^{-i\omega t+i{\bm k}\cdot{\bm r}}
+a_R^\dagger({\bm k})e^{i\omega t-i{\bm k}\cdot{\bm r}}\right),
\end{align}
where ${\bm e}({\bm k})={\bm e}_L({\bm k})$ is the normalized complex polarization vector corresponding to the left-handed circular polarization (positive helicity). The vector ${\bm e}^*({\bm k})={\bm e}_R({\bm k})$ describes the right-handed polarization. The operator $a_L({\bm k})$ annihilates the left-handed photon with the wave vector ${\bm k}$, and $a_R^\dagger({\bm k})$ creates the right-handed photon. The interaction Hamiltonian (\ref{hint}) expressed in terms of ${\bm F}$ reads
\begin{equation}\label{ham}
H_I=\frac{\alpha}{4}\left({\bm F}+{\bm F}^\dagger\right)^2-\frac{\beta}{4}\left({\bm F}-{\bm F}^\dagger\right)^2.
\end{equation}
Anticipating the smallness of the whole effect, we shall consider only the lowest order of perturbation theory:
\begin{equation}
S\approx 1 - \frac{i}{\hbar}\int\!d^3r\,dt\,H_I({\bm r}, t).
\end{equation}

\section{Radiation produced by orbiting dielectric or magnetic spheres}

As an introduction to the main subject, we consider the electromagnetic radiation produced by a homogeneous dielectric or magnetic sphere in a uniform motion on a circular orbit. In the case of a dielectric sphere of radius $a$, the function $\alpha({\bm r}, t)$ can be written in the form
\begin{equation}
\alpha({\bm r}, t)=\kappa g({\bm r}-{\bm R}(t)),
\end{equation}
where $g({\bm r})=\theta(a-r)$ refers to the sphere at rest, ${\bm R}(t)=R(\cos(\Omega t),\sin(\Omega t),0)$  describes the orbital motion, $\kappa=(1/\epsilon_r-1)$, and $\epsilon_r$ is the relative static dielectric constant. For a magnetic sphere, the only difference is that in all the formulas $\epsilon_r$ is to be replaced by $\mu_r$.

Assuming that initially there are no photons, the relevant terms of $H_I$ are only those containing the products of two creation operators, i.e.,
\begin{align}\label{s}
S|0\rangle &\approx |0\rangle
-\frac{i}{4(2\pi)^3}\sum_{\lambda_1,\lambda_2}\int\!d^3k_1\int\!d^3k_2\nonumber\\
&\times\sqrt{\omega_1\omega_2}\,
{\tilde\alpha}({\bm k}_1+{\bm k}_2,\omega_1+\omega_2)\nonumber\\
&\times{\bm e}_{\lambda_1}^*({\bm k}_1)\!\cdot\!{\bm e}_{\lambda_2}^*({\bm k}_2)a_{\lambda_1}^\dagger({\bm k}_1)a_{\lambda_2}^\dagger({\bm k}_2)|0\rangle,
\end{align}
where ${\tilde\alpha}({\bm k},\omega)$ is the four-dimensional Fourier transform of $\alpha({\bm r}, t)$,
\begin{align}\label{delta}
{\tilde\alpha}({\bm k},\omega)&=\int\!d^3re^{-i{\bm k}\cdot{\bm r}}g({\bm r})\int\!dte^{i\omega t}e^{i{\bm k}\cdot{\bm R}(t)}\nonumber\\
&=\!\!\sum_{m=-\infty}^{\infty}\!\!{\tilde\alpha}_m({\bm k},\omega)2\pi\delta(\omega-m\Omega),\\
{\tilde\alpha}_m({\bm k},\omega)&=\frac{4\pi a^3}{3}\kappa\,i^me^{im\varphi} f(ka)J_m(k_\perp R),
\end{align}
$k_\perp$ and $\varphi$ are the polar coordinates of ${\bm k}$ in the orbital plane, $k=|{\bm k}|$,  and
\begin{align}
f(ka)=\frac{3}{(ka)^3}\left[\sin(ka)-ka\cos(ka)\right]\approx1-\frac{(k a)^2}{10}.
\end{align}
The decomposition into Bessel functions in order to identify the contributions with different frequencies has been used before by Nienhuis \cite{nien} in his study of a rotating lens.

According to formula (\ref{s}), an orbiting dielectric sphere radiates pairs of photons. Note, that the Planck constant does not appear in Eq.~(\ref{s}), even though the process of photon pair creation is {\em a purely quantum-mechanical effect}. This effect is not predicted by classical theory. The Planck constant will reappear in the formula for the radiated power.

It follows from Eq.~(\ref{delta}) that the sum of photon energies is a multiple of $\hbar\Omega$. Therefore, the sum over $m$ is effectively restricted to positive values of $m$. The probability $w_{\lambda_1\lambda_2}({\bm k}_1,{\bm k}_2)$ for emitting a pair of photons per unit time is
\begin{align}
&w_{\lambda_1\lambda_2}({\bm k}_1,{\bm k}_2)=\frac{1}{(2\pi)^5}\frac{|{\bm e}_{\lambda_1}({\bm k}_1)\!\cdot\!{\bm e}_{\lambda_2}({\bm k}_2)|^2}{8}\nonumber\\
&\times\sum_{m=1}^\infty\!\omega_1\omega_2|{\tilde\alpha}_m({\bm K},\varpi)|^2 \delta(\varpi-m\Omega)\\
&=\frac{1}{144\pi^3}
\frac{(1\mp\cos\theta)^2}{4}\nonumber\\
&\times\sum_{m=1}^\infty\!\omega_1\omega_2\kappa^2a^6f^2(Ka)J^2_m(K_\perp R)\delta(\varpi-m\Omega),\label{bessel}
\end{align}
where ${\bm K}={\bm k}_1+{\bm k}_2$, $\varpi=\omega_1+\omega_2$, and ${\bm K}_\perp={\bm k}_{\perp 1}+{\bm k}_{\perp 2}$. We employed here the standard trick $2\pi\delta(\omega-\omega')^2\to T\delta(\omega-\omega')$, and we also used the relation
\begin{align}
|{\bm e}_{\lambda_1}({\bm k}_1)\!\cdot\!{\bm e}_{\lambda_2}({\bm k}_2)|^2=(1\mp\cos\theta)^2/4.
\end{align}
The minus (plus) sign should be taken for the same (opposite) polarizations and $\theta$ is the angle between the vectors ${\bm k}_1$ and ${\bm k}_2$. The presence of $1\mp\cos\theta$ shows that photons with the same polarization predominantly travel in opposite directions whereas photons with opposite polarizations travel together. However, the probability of photons traveling in opposite directions is suppressed due to the small values of the Bessel functions near zero. Moreover, the factor $\omega_1\omega_2$ reaches its maximum when $\omega_1=\omega_2$, since $\omega_1+\omega_2$ is fixed. Therefore, the most likely phenomenon is the emission of a photon pair with equal wave vectors and opposite circular polarizations.

The sum over all Bessel functions represents contributions from pairs of photons whose total energy is $m\hbar\Omega$. However, the intensities of higher harmonics are strongly suppressed. In order to quantify this statement, let us note that the arguments of the Bessel functions in Eq.~(\ref{bessel}) cannot exceed $mR\Omega/c=mv_R$, where
\begin{align}
v_R=R\Omega/c
\end{align}
is the orbital velocity measured relative to the speed of light. Since for small $v_R$ the ratio of two consecutive terms is approximately equal to
\begin{align}
\frac{(m+1)^2J_{m+1}^2((m+1)v_R)}{m^2J_m^2(mv_R)}\approx v_R^2\left(1+\frac{1}{m}\right)^{2m+2},
\end{align}
the contributions from higher harmonics can all be neglected and we can keep only the ${\tilde\alpha}_1$ term.

The angular distribution of the emitted pairs is described by the Bessel function $J_1^2(KR\cos\chi)$, where $\chi$ is the angle between the direction of the photon pair and the orbital plane. Since $KR$ is small, the angular distribution is governed by $\cos^2\chi$. Therefore, the photons are emitted predominantly in the orbital plane. No photons are emitted in the direction normal to the plane.

The main contribution to the total rate for emitting a pair of photons is
\begin{align}\label{total}
&w=\frac{\kappa^2a^6}{144\pi^3}\!\!\int\!d^3k_1d^3k_2(1+\cos^2\theta)\nonumber\\
&\times\omega_1\omega_2f^2(Ka) J^2_1(K_\perp R)\delta(\varpi-\Omega).
\end{align}
To obtain an order-of-magnitude estimate of $w$, we introduce dimensionless photon wave vectors ${\bm l}_1=c{\bm k}_1/\Omega$ and ${\bm l}_2=c{\bm k}_2/\Omega$:
\begin{align}\label{total1}
&w=\frac{\kappa^2}{72\pi^2T}\left(\frac{a}{R}\right)^6v_R^6\int\!d^3l_1d^3l_2(1+\cos^2\theta)\nonumber\\
&\times l_1 l_2 f^2(Lv_Ra/R) J^2_1(L_\perp v_R)\delta(l_1+l_2-1),
\end{align}
where $T$ is the orbital period. For a nonrelativistic orbital velocity $f^2(Lv_Ra/R)\approx1$ and $J^2_1(L_\perp v_R)\approx (L_\perp v_R)^2/4$. Hence, the total rate can be written in the form
\begin{align}\label{total2}
&w\approx\frac{\kappa^2}{288\pi^2T}\left(\frac{a}{R}\right)^6v_R^8\nonumber\\
&\times\int\!d^3l_1d^3l_2(1+\cos^2\theta) l_1 l_2  L_\perp^2\delta(l_1+l_2-1).
\end{align}
The integral in second line is a pure number of the order of 1. Therefore, the total power of emitted radiation is approximately equal to
\begin{align}\label{power}
P_E\approx\frac{\kappa^2}{288\pi^2}\left(\frac{a}{R}\right)^6v_R^8\,\frac{\hbar\Omega}{T}.
\end{align}
For a nonrelativistic motion, this value is extremely small, as was to be expected. It is worth noting that the probability of photon emission by a magnetic orbiting sphere differs only in having $\kappa^2=(1/\epsilon_r-1)^2$ replaced by $(1/\mu_r-1)^2$.

To compare the photon production by the orbiting objects, caused exclusively by their electromagnetic properties, with a similar effect caused by a time-varying gravitational field, we shall calculate now the radiation by two dielectric spheres orbiting around their center of mass. The radiation of this system due to the time-dependent gravitational field will be discussed in the next section. We assume that the bodies with masses $M_1=\mu M$ and $M_2=(1-\mu) M$ and radii $a_1$ and $a_2$ move on circular orbits with frequency $\Omega$. Their trajectories are given then by the formulas
\begin{subequations}
\begin{align}\label{traj}
{\bm R}_1(t)&=(1-\mu) R(\cos(\Omega t),\sin(\Omega t),0),\\
{\bm R}_2(t)&=-\mu R(\cos(\Omega t),\sin(\Omega t),0),
\end{align}
\end{subequations}
where $R=(G M/\Omega^2)^{1/3}$. In this case,
\begin{align}\label{alpha2}
&{\tilde\alpha}_m({\bm k},\omega)=i^me^{im\varphi}M\big[\mu\frac{\kappa_1}{\rho_1} f(ka_1)J_m((1-\mu)k_\perp R)\nonumber\\
&+(-1)^m(1-\mu)\frac{\kappa_2}{\rho_2} f(ka_2)J_m(\mu k_\perp R)\big],
\end{align}
where $\rho_1$ and $\rho_2$ are the average densities. The dominant contribution still comes from ${\tilde\alpha}_1$, unless the material coefficients are fine-tuned so that $\kappa_1/\rho_1=\kappa_2/\rho_2$. In all the remaining cases, we may replace the difference $\kappa_1/\rho_1-\kappa_2/\rho_2$ by some effective value ${\bar\kappa}/{\bar\rho}$:
\begin{align}\label{alpha3}
|{\tilde\alpha}_1({\bm k},\omega)|^2\approx \frac{M^2k_\perp^2R^2\mu^2(1-\mu)^2}{4}
\left(\frac{\bar\kappa}{\bar\rho}\right)^2.
\end{align}
The counterpart of formula (\ref{total2}) has the form
\begin{align}\label{total3}
&w\approx\frac{M^2\mu^2(1-\mu)^2}{512\pi^4T}
\left(\frac{\bar\kappa}{\bar\rho}\right)^2\frac{v_R^8}{R^6}\nonumber\\
&\times\int\!d^3l_1d^3l_2(1+\cos^2\theta) l_1 l_2  L_\perp^2\delta(l_1+l_2-1).
\end{align}
Therefore, the radiated power is approximately equal to
\begin{align}\label{power2}
&P_E\approx\frac{M^2\mu^2(1-\mu)^2}{512\pi^4}\left(\frac{\bar\kappa}{\bar\rho}\right)^2\frac{v_R^8}{R^6}\frac{\hbar\Omega}{T}.
\end{align}

In the special case, when $\kappa_1/\rho_1=\kappa_2/\rho_2=\kappa/\rho$, the leading term in the contribution from $m=1$ vanishes and the contribution from $m=2$ becomes dominant,
\begin{align}\label{alpha4}
|{\tilde\alpha}_2({\bm k},\omega)|^2\approx\frac{M^2k_\perp^4R^4\mu^2(1-\mu)^2}{64}\left(\frac{\kappa}{\rho}\right)^2,
\end{align}
and the radiated power, as compared to the generic case (\ref{power2}), decreases roughly by a factor of $v_R^2$.

\section{Radiation produced by a time-varying metric}

A double-star system creates a time-dependent gravitational field. We shall consider an isolated double-star system that modifies the metric only in its neighborhood. The electromagnetic radiation produced by such a system will be analyzed from the point of view of an observer located far away, where the space is flat.

Of course, the exact form of the metric field for a two-body system is not known. However, for a ordinary stars (even for neutron stars) the Schwarzschild radius $r_0$ is much smaller than the geometric radius and we can use the weak-field approximation. In this approximation the diagonal components $h_0$ and $h_1$ of the exact Schwarzschild metric in isotropic coordinates (cf., for example, \cite{dinv})
\begin{align}
ds^2=h_0c^2dt^2-h_1(dx^2+dy^2+dz^2),\label{metric}\\
h_0=\left(\frac{1-r_0/4r}{1+r_0/4r}\right)^2,\;\;h_1=\left(1+\frac{r_0}{4r}\right)^4,
\end{align}
can be replaced by
\begin{align}
h_0\approx 1-\frac{r_0}{r},\;\;h_1\approx 1+\frac{r_0}{r}.
\end{align}
In this approximation, the theory is linearized and we can simply add the contributions to the metric from both bodies. For two orbiting stars, the metric components outside the stars are
\begin{align}
h_0({\bm r},t)&=1-\frac{r_1}{|{\bm r}-{\bm R}_1(t)|}-\frac{r_2}{|{\bm r}-{\bm R}_2(t)|},\\
h_1({\bm r},t)&=1+\frac{r_1}{|{\bm r}-{\bm R}_1(t)|}+\frac{r_2}{|{\bm r}-{\bm R}_2(t)|},
\end{align}
where $r_1$ and $r_2$ are the Schwarzschild radii of the stars:
\begin{align}
r_1=\mu\frac{2GM}{c^2},\;\;r_2=(1-\mu)\frac{2GM}{c^2}.
\end{align}

We have chosen a particular form of the metric tensor, but the calculated photon production rates are invariant under a change of the gravitational gauge. In the linearized version of gravity this change is given by (cf., for example, \cite{wald})
\begin{align}\label{ggauge}
\delta g_{\mu\nu}(x)=\partial_\mu\xi_\nu(x)+\partial_\nu\xi_\mu(x).
\end{align}
The proof of the invariance of the $S$ operator under this transformation is given in the Appendix.

The Maxwell equations derived from the variational principle based on the Lagrangian
\begin{align}
{\cal L} = -\frac{\epsilon_0}{4}\sqrt{-g}g^{\mu\lambda}g^{\nu\rho}f_{\mu\nu}f_{\lambda\rho},
\end{align}
written for an arbitrary metric, have the same form (\ref{max}), but the constitutive relations are modified. In our case, for the diagonal metric, these relations are (cf., for example, \cite{ll})
\begin{align}
{\bm E}({\bm r}, t)=\frac{1}{\epsilon_0}\sqrt{\frac{h_0({\bm r}, t)}{h_1({\bm r}, t)}}{\bm D}({\bm r}, t),\\
{\bm H}({\bm r}, t)=\frac{1}{\mu_0}\sqrt{\frac{h_0({\bm r}, t)}{h_1({\bm r}, t)}}{\bm B}({\bm r}, t).
\end{align}
Therefore,
\begin{align}
\alpha({\bm r}, t)&=\beta({\bm r}, t)=\sqrt{\frac{h_0({\bm r}, t)}{h_1({\bm r}, t)}}-1\\
&\approx-\frac{r_1}{|{\bm r}-{\bm R}_1(t)|}-\frac{r_2}{|{\bm r}-{\bm R}_2(t)|}.
\end{align}

The interaction Hamiltonian is now equal to
\begin{equation}\label{ham1}
H_I=\alpha{\bm F}\!\cdot\!{\bm F}^\dagger,
\end{equation}
and the counterpart of formula (\ref{s}) is
\begin{align}\label{s1}
S|0\rangle &\approx |0\rangle
-\frac{i}{(2\pi)^3}\int\!d^3k_1\int\!d^3k_2\nonumber\\
&\times\sqrt{\omega_1\omega_2}\,{\tilde\alpha}({\bm k}_1+{\bm k}_2,\omega_1+\omega_2)\nonumber\\
&\times{\bm e}_L^*({\bm k}_1)\!\cdot\!{\bm e}_R^*({\bm k}_2)a_L^\dagger({\bm k}_1)a_R^\dagger({\bm k}_2)|0\rangle.
\end{align}
Thus, the emitted photons always have opposite circular polarizations.
The Fourier transform of $\alpha({\bm r},t)$ is
\begin{align}\label{alpha}
&{\tilde\alpha}({\bm k},\omega)=-\frac{2GM}{c^2}
\big(\mu\int\limits_{r>a_1}\!\!\!\frac{d^3r}{r}e^{i{\bm k}\cdot{\bm r}}\int\limits_{-\infty}^{\;\infty}\!\!\!dt\, e^{i\omega t}e^{i(1-\mu){\bm k}\cdot{\bm R}(t)}\nonumber\\
&+(1-\mu)\int\limits_{r>a_2}\!\!\!\frac{d^3r}{r}e^{i{\bm k}\cdot{\bm r}}\int\limits_{-\infty}^{\;\infty}\!\!\!dt\, e^{i\omega t}e^{-i\mu{\bm k}\cdot{\bm R}(t)}\big).
\end{align}
Its decomposition into harmonics [cf. Eq.~(\ref{delta})] gives
\begin{align}\label{alpham}
&{\tilde\alpha}_m({\bm k},\omega)=-\frac{16\pi^2}{k^2}
i^m e^{im\varphi}\Big[r_1\cos(ka_1)J_m((1-\mu)k_\perp R)\nonumber\\
&+(-1)^mr_2\cos(ka_2)J_m(\mu k_\perp R)\Big].
\end{align}

The probability of a pair production per unit time is
\begin{align}
&w({\bm k}_1,{\bm k}_2)=\frac{2\omega_1\omega_2}{(2\pi)^5}|{\bm e}_R({\bm k}_1)\!\cdot\!{\bm e}_L({\bm k}_2)|^2\nonumber\\
&\times\sum_{m=1}^\infty\!|{\tilde\alpha}_m({\bm K},\varpi)|^2 \delta(\varpi-m\Omega).
\end{align}
As in the previous case, the photons are predominantly emitted in the orbital plane with the same wave vectors.

In the present case of a nonrelativistic motion of two orbiting bodies, due to a cancellation of the leading terms, $|{\tilde\alpha}_1({\bm K},\varpi)|^2$ can be neglected as compared to $|{\tilde\alpha}_2(2{\bm K},2\varpi)|^2$. The ratio
\begin{align}
\frac{|{\tilde\alpha}_1({\bm K},\varpi)|^2}{|{\tilde\alpha}_2(2{\bm K},2\varpi)|^2}&\approx v_R^2
\left(\frac{1-2\mu}{2}+2\frac{a_1^2-a_2^2}{R^2}\frac{K}{K_\perp}\right)^2
\end{align}
is small, of the order of $v_R^2$, except for $K_\perp\ll K$, but this region contributes very little to the total rate.

Keeping only the $m=2$ term, we obtain the following final approximate formula for the total rate (after the substitutions ${\bm k}_1=2\Omega{\bm l}_1/c$ and ${\bm k}_2=2\Omega{\bm l}_2/c$)
\begin{align}
&w\approx \frac{64}{\pi^2T}\mu^2(1-\mu)^2 v_R^{10}\nonumber\\
&\times\int\!d^3l_1d^3l_2(1+\cos\theta)^2\frac{l_1 l_2 L_\perp^4}{L^4}\delta(l_1+l_2-1),
\end{align}
where we have used the relation $(GM\Omega/c^3)=v_R^3$. The integral, similar to the one in Eq.~(\ref{total2}), is a pure number of the order of 1. Therefore, the total power radiated by gravitating stars caused by the time-varying metric is
\begin{align}\label{power0}
P_M\approx\frac{64}{\pi^2}\mu^2(1-\mu)^2v_R^{10}\,\frac{2\hbar\Omega}{T} .
\end{align}
For a double star of two solar masses and orbiting period of 1 h, we obtain  $v_R=0.0026$ and therefore
\begin{align}
P_M\approx5.4\times 10^{-27}\,\frac{2\hbar\Omega}{T}.
\end{align}
This effect is exceedingly small --- one has to wait more than $10^{22}$ yr for a pair of photons to be emitted. For solar-type stars the gravitational effect is weaker by three orders of magnitude than the effect due to the time-dependence of the dielectric constant. However, the radiation due to a disturbance of the metric might be stronger than the radiation due to the electric (or magnetic) properties of the same double-star system. The ratio of the radiated power due to the gravitational effects (\ref{power0}) to that due to the electromagnetic effects (\ref{power2}) is
\begin{align}\label{ratio}
\frac{P_M}{P_E}\approx 6.5\times 10^5\left(\frac{\bar\rho}{\bar\kappa}\right)^2 \frac{G^{8/3}M^{2/3}}{c^2\Omega^{10/3}}.
\end{align}
This ratio varies quadratically with the density. When one of the stars is replaced by a typical neutron star with the density of about $10^{17}{\rm kg/m^3}$, the gravitational effects dominate over the electromagnetic effects by 25 orders of magnitude.

It is also worth stressing that the orbital velocity appears in Eq.~(\ref{power0}) to the tenth power. Therefore the effect would increase dramatically in the relativistic case. A significant increase of the velocity will change the power of emitted radiation by many orders of magnitude.

In order to complete this investigation we shall compare it now with the power radiated by the same system in the form of gravitons. The emission of gravitons, like the emission of photons by an external current, is a purely classical phenomenon and we can expect that the associated power $P_G$ will be much larger. Indeed, the formula for $P_G$ in the case of two gravitating bodies moving on circular orbits reads in our notation (cf., for example, \cite{ll,hartle})
\begin{align}\label{pg}
P_G = \frac{64\pi}{5}v_R^7\frac{Mc^2}{T}.
\end{align}
Thus, the power radiated by double stars in the form of gravitons is enormous as compared to the power radiated in the form of photons due to the time-varying metric.

In principle, at the more fundamental level, the production of photon pairs by gravitating bodies could be, in our opinion, described as a two-step process. In the first step, as predicted by linearized quantum gravity, the system produces a lot of gravitons. In the second step, colliding gravitons produce pairs of photons (a single graviton cannot decay) due to a coupling of gravitons to photons [described by the interaction Hamiltonian (\ref{ham1})]. We believe that this mechanism will give results equivalent to those obtained by our semiclassical treatment of gravitational effects. However, the calculation of the photon production rates along these lines would be much more complicated.

\appendix
\section{}

Our proof of the gravitational gauge invariance is patterned after a proof of the electromagnetic gauge invariance in quantum electrodynamics. In QED a change of the $S$ operator under a change $\delta{\cal A}_\mu$ of the external electromagnetic field is
\begin{align}\label{elg}
\delta S=-iT\int\!d^4x\, \left[j^\mu(x)S\right] \delta{\cal A}_\mu(x).
\end{align}
This fundamental formula may even be viewed as a definition of the four-current through the linear response of the system to a change of the external field. When the variation $\delta{\cal A}_\mu$ is only a change of gauge,
\begin{align}\label{elg1}
\delta{\cal A}_\mu(x)=\partial_\mu\Lambda,
\end{align}
the integral in Eq.~(\ref{elg}), after integration by parts, vanishes due to current conservation, $\partial_\mu j^\mu=0$.

In the case of gravity, the formula describing the change of $S$ under a change of the metric tensor, a counterpart of Eq.~(\ref{elg}), reads
\begin{align}\label{gravg}
\delta S=-\frac{i}{2}{\rm T}\int\!d^4x\, \left[T^{\mu\nu}(x)S\right] \delta g_{\mu\nu}(x),
\end{align}
where $T^{\mu\nu}$ is the energy-momentum tensor describing the coupling of matter to gravity. Inserting into this equation formula (\ref{ggauge}) for the change of the metric tensor due to a change of the coordinates, we obtain after integration by parts
\begin{align}\label{gravg1}
\delta S=iT\int\!d^4x\, \partial_\mu \left[T^{\mu\nu}(x)S\right] \xi_{\mu}(x).
\end{align}
This integral vanishes because in linearized gravity $\partial_\mu T^{\mu\nu}=0$. In our proof we argue that because the energy-momentum tensor is conserved, the $S$ operator is a scalar, invariant under the transformations of the coordinate system. This is the reverse path to that followed often in general relativity (cf., for example, \cite{wein}), where it is argued that the energy-momentum tensor is conserved because the action is a scalar.

\end{document}